\documentclass[aps,prb,reprint,floatfix,superscriptaddress]{revtex4-1}

\usepackage[english]{babel} 
\usepackage{hyperref}
\usepackage{amssymb,amsmath,upgreek}
\usepackage{graphicx}
\usepackage{float}
\usepackage{url}
\usepackage{multirow}
\usepackage{chemformula}
\usepackage{tabularx}
\usepackage{placeins}
\usepackage{mathtools}

\usepackage[utf8]{inputenc}

\newlength{\smallpic}
\begin{document}
\newcommand{\dprime}{{\prime \prime}}

\title{Critical temperatures of two dimensional magnets beyond linear spin wave theory: application to CrI$_3$, MPS$_3$ (M=Ni, Mn, Fe) and CrSBr}

\author{Varun Rajeev Pavizhakumari}
\email{Corresponding author: vrapa@dtu.dk, \\ varun.rp3@gmail.com}
\affiliation{Computational Atomic-Scale Materials Design (CAMD), Department of Physics, Technical University of Denmark, 2800 Kgs. Lyngby, Denmark}
\author{Thomas Olsen}
\affiliation{Computational Atomic-Scale Materials Design (CAMD), Department of Physics, Technical University of Denmark, 2800 Kgs. Lyngby, Denmark}

\begin{abstract}
Magnetic anisotropy is crucial for sustaining long range magnetic order in two-dimensional materials (2D) and must be taken into account by any approximate scheme for calculating critical temperatures. 
Despite the huge interest in two-dimensional magnetic materials, a general and reliable method for calculating the critical temperature is still lacking.
The concept of 
ferromagnetism and 
anti-ferromagnetism can be regarded as special cases of single-$Q$ magnetic order, and for such systems the critical temperature can be calculated from the magnon dispersion using either Holstein-Primakoff (HP) bosonization or Green's function-based Random Phase Approximation (RPA). Here, we study the effects of single-ion anisotropy in general single-$Q$ systems in both the HP and RPA methods. In the case of RPA, we generalize the approach to include the Callen Decoupling (CD) correction, which has previously been shown to yield good agreement with experimental Curie temperatures for 2D ferromagnets. 
We compare the calculated critical temperatures of CrI$_3$ (uniaxial ferromagnet), MPS$_3$ (M=Ni, Mn, Fe) (uniaxial anti-ferromagnets) and CrSBr (triaxial ferromagnet) monolayers with experimental values and find that the Green's function-based methods are much more reliable than HP and that the CD decoupling appears to be more accurate than RPA if the single-ion anisotropy is large.
\end{abstract}

\maketitle

\section{Introduction}
Since the discovery of ferromagnetic order in monolayer CrI$_3$ \cite{Huang2017} there has been a increasing interest in magnetic two-dimensional materials. In addition to the fundamental physical aspects associated with two-dimensional (2D) magnetism, a wide variety of possible spintronics applications have been proposed \cite{Burch2018,Wang2022-rev}. Compared to three-dimensional (3D) magnetic materials, a major difference is the crucial role played by magnetic anisotropy. The Mermin-Wagner (MW) theorem states that continuous symmetries cannot be {\it spontaneously} broken at finite temperature in 2D systems with short-range interactions \cite{MW1966} and this implies that order requires {\it explicit} symmetry breaking by magnetic anisotropy. 
While one may still obtain a reliable description of the magnetic interactions in terms of  Heisenberg models, well-established methodology for calculating magnetic properties of such models may break down in 2D. For example, conventional Weiss mean field theory predicts a finite critical temperature in any isotropic model - irrespective of the dimensionality of the lattice - and thus fails rather dramatically in 2D. 
On the other hand, correlation effects can be accounted for by applying classical Monte Carlo (MC) simulations, which are commonly used to estimate critical temperatures \cite{Torelli2019,Strandburg-mc1992,Xiaobo-mc2019,Xue-mc2022}. However, such calculations are often computationally demanding and  accuracy is typically limited by the classical treatment \cite{Helimag3d2025}. Finally, one may try to obtain the thermal properties directly from the quantum mechanical excitations, which can be calculated by various approximate methods. In the case of long range order, the fundamental magnetic excitations of the Heisenberg model will be comprised by magnons, and while single-magnon energies are easily calculated, magnon-magnon interactions have to be approximated at the mean-field level. This leads to a temperature-dependent renormalization of magnon energies, which can be calculated in either the Holstein Primakoff (HP) bosonization scheme or in the Green's function-based Random Phase Approximation (RPA) \cite{HP1940,tyablikov1959,Tahir1962}. These methods have recently been developed and benchmarked against experimental critical temperatures in 3D materials \cite{Helimag3d2025}, but due to the importance of magnetic anisotropy in 2D, the application to 2D magnetism calls for a study of its own.

At the level of two-spin interactions, spin-orbit coupling may give rise to either anisotropic exchange (two-site) interactions or single-ion (one-site) anisotropy. 
The former case can be handled rather generally by introducing an exchange tensor describing the magnetic interaction between two sites and is straightforward to include in both HP \cite{Toth2015} and RPA \cite{Helimag3d2025}. The single-ion anisotropy may be included in these approaches as well, but the associated mean-field approximation becomes much more subtle because the two spin operators do not commute at the same site. 
For example, the generalized single-$Q$ method of Ref. \cite{Helimag3d2025}, can handle uniaxial, biaxial and triaxial single-ion anisotropies in the Heisenberg model, but RPA fails to predict the irrelevance of single-ion anisotropy for the case of $S=1/2$ systems. The HP method (with a mean-field treatment of two-magnon interactions) correctly captures the vanishing spin-wave gap 
(irrelevance of single-ion anisotropy) for uniaxial $S=1/2$ ferromagnets at $T=0$, but fails to do so at elevated temperatures as well as tri-axial systems at $T=0$. 
An alternative to RPA was proposed by Herbert Callen in 1963 \cite{Callen1963} and will be referred to as Callen Decoupling (CD) in the following. Callen's scheme comprises an alternative mean-field decoupling of the equation of motion, which may be applied to the calculations of critical temperatures. We have recently shown that in this respect the method performs better than RPA in the classical limit (relevant for itinerant systems), but not in the quantum regime \cite{BeyondRPA2025}.  
However, CD gives the correct behavior for the single-ion anisotropy contribution to the magnon dispersion (irrelevance of single-ion anisotropy and vanishing spin-wave gap for $S=1/2$ systems), and it has been shown that augmenting the RPA method for exchange interactions with CD for the single-ion anisotropy (RPA+CD) gives excellent agreement with exact results for ferromagnets \cite{BeyondRPA2025,Frobich2006}. The CD has been developed for simple anti-ferromagnets \cite{Callen1964-afm,Liu-CD1967} as well, but a general approach for arbitrary single-$Q$ states seems to be lacking in the literature.

In this work, we conduct a systematic study of the effect of single-ion anisotropy on the prediction of critical temperature obtained by various approximations. The methods are applied to well-known 2D magnets that represent different types of order: CrI$_3$ (uniaxial ferromagnet), MPS$_3$ (M=Ni, Mn, Fe) (uniaxial anti-ferromagnets) and CrSBr (triaxial ferromagnet).  We use the experimental exchange constants obtained from inelastic neutron scattering experiments and compare the critical temperatures calculated with HP, RPA, RPA+CD and classical MC with experimental values. 

The paper is organized as follows. In sec. \ref{sec:HP}, we generalize the HP method to include a mean-field description of magnon-magnon interactions for arbitrary single-$Q$ states. In sec. \ref{sec:Gfn}, we include the single-ion anisotropy tensor in both the RPA and the CD approach for single-$Q$ ordered systems. Sec. \ref{sec:limitingcases}, details the behavior of 2D critical temperatures as the single-ion anisotropy approaches zero and Sec. \ref{sec:dipole} generalizes the theory  to include  dipole-dipole interactions. Finally, in sec. \ref{sec:results}, we calculate the critical temperature of CrI$_3$, MPS$_3$ (M=Ni, Mn, Fe) and CrSBr and compare the performance of the different methods.

\section{Theory}\label{sec:theory}
Our starting point is the Heisenberg model
\begin{equation}\label{eq:heisenberg-model}
    \mathcal{H} = -\frac{1}{2}\sum_{abij} \mathbf{S}^T_{ai}\mathrm{J}_{abij}\mathbf{S}_{bj}
\end{equation}
where $\mathbf{S}_{ai}$ is the spin operator for site $a$ in unit cell $i$ and $\mathrm{J}_{abij}$ is the exchange tensor that couples spins at $i,a$ and $j,b$. 
We will denote the number of unit cells by $N$ and the number of magnetic sites per unit cell by $N_a$.
Written in this form the model includes single-ion anisotropy represented by the terms with $a=b$ and $i=j$. In the following it will be practical to consider such terms separately and we write them as
\begin{equation}
    \mathcal{H}_\mathrm{K} = - \sum_{ai} \mathbf{S}^T_{ai}\mathrm{K}_{ai}\mathbf{S}_{ai} \label{eq:heisenberg-model_sia},
\end{equation}
with $\mathrm{K}_{ai}=\mathrm{J}_{aaii}/2$ being the single-ion anisotropy tensor. It should be noted that the magnon dispersion must be invariant under the addition of an isotropic term to the single-ion anisotropy tensor: $\mathrm{K}_{ai}^{\mu\nu}\rightarrow \mathrm{K}_{ai}^{\mu\nu}+\alpha\delta^{\mu\nu}$. This is due to the fact that $\mathbf{S}_{ai}\cdot\mathbf{S}_{ai}$ is a constant and will be referred to as isotropic invariance. The freedom will be utilized to enforce certain components of the tensor to vanish in the following. It is also clear that the single-ion anisotropy term becomes a constant (and thus irrelevant) for any $S=1/2$ system.

In the present work we focus on a rather general type of magnetic order referred to as single-$Q$ states. For such cases the ground state magnetic structure is characterized by an ordering vector $\mathbf{Q}$, a global rotation axis $\mathbf{\hat{n}}$ and the relative direction of the magnetic moments in a particular unit cell. The spins in different unit cells are then related by $\langle\mathbf{S}_{ai}\rangle=\mathrm{R}_\mathbf{\hat{n}}(\mathbf{Q}\cdot \mathbf{r}_{ij})\langle\mathbf{S}_{aj}\rangle$, where $\mathrm{R}_\mathbf{\hat{n}}$ is a rotation matrix and $\mathbf{r}_{ij}$ is a lattice vector connecting cell $i$ to cell $j$. It will be convenient to transform the system to a local coordinate system where all spin fluctuations are perpendicular to the local-$z$ axis. 
This entails a rotation of all spin operators into a locally ferromagnetic system and can be divided into two steps: a cell-dependent rotation $\mathrm{R}_i$ and a site-dependent (cell-independent) rotation $\mathrm{R}^\prime_a$. The spin operators are then written as
\begin{align}
     \mathbf{S}_{ai} &= \mathrm{R}_i \mathbf{S}'_{ai} \label{eq:rot-uc},\\
     \mathrm{R}_i &= \mathrm{R}_\mathbf{\hat{n}}(\mathbf{Q}\cdot \mathbf{r}_i),\\
     \mathbf{S}^\prime_{ai} &= \mathrm{R}^\prime_a \mathbf{S}^\dprime_{ai}, \label{eq:rot-site}
\end{align}
where $\hat{\mathbf{n}}$ is the global rotation axis for the single-$Q$ spiral state, $\mathbf{S}'_{ai}$ are spins in the coordinates that are independent of unit cells and $\mathbf{S}^\dprime_{ai}$ are spins in the locally ferromagnetic frame independent of sites \cite{Toth2015,Helimag3d2025}. Introducing two new vectors $\mathbf{u}_a$ and $\mathbf{v}_a$
\begin{align}
    [\mathbf{u}_a]^\mu &= \big[{\mathrm{R}^\prime_a}\big]^{\mu x} + i\big[{\mathrm{R}^\prime_a}\big]^{\mu y} \label{eq:u_a},\\
    [\mathbf{v}_a]^\mu &= \big[{\mathrm{R}^\prime_a}\big]^{\mu z}\label{eq:v_a}.
\end{align}
Eq. \eqref{eq:heisenberg-model_sia} can be written as
\begin{align}
    \mathcal{H}_\mathrm{K} &= -\sum_{ai} \mathbf{S}^{\prime T}_{ai}  \mathrm{K}_{a}\mathbf{S}^{\prime}_{ai}\label{eq:rot-heisenberg-model_sia},
\end{align}
where
\begin{align}\label{eq:Sprime-uv}
    \mathbf{S}^{\prime}_{ai} = \mathbf{u}_a \frac{S^{\dprime +}_{ai}}{2} + \mathbf{u}^*_a \frac{S^{\dprime -}_{ai}}{2} + \mathbf{v}_a S^{\dprime z}_{ai}.
\end{align}
and we used the rotational symmetry of the anisotropy tensor, which implies $\mathrm{R}^{T}_{i}\mathrm{K}_{ai} \mathrm{R}_{i}=\mathrm{K}_{ai}=\mathrm{K}_a$. 

Before we proceed we note that the presence of single-ion anisotropy may or may not be consistent with the assumption of a single-$Q$ state. For example, the well-known case of a 2D hexagonal lattice with nearest neighbor anti-ferromagnetic isotropic interactions, gives rise to a ground state with $120^\circ$ between all spins. This is a planar spin spiral, which can be described as a single-$Q$ state with $\mathbf{Q}=(1/3,1/3)$. If single-ion anisotropy is included, symmetry dictates that it must be  uniaxial, which implies an easy-plane or an easy-axis. In the former case, the spiral plane coincides with the lattice plane and the single-$Q$ structure is preserved. However, for easy-axis anisotropy the spiral plane becomes orthogonal to the lattice and the $120^\circ$ symmetry will be slightly (if anisotropy is small) broken. The compatibility of a given single-$Q$ state (including orientation) with a single-ion anisotropy tensor is essentially captured by whether or not the condition $\mathrm{R}^{T}_{i}\mathrm{K}_{ai} \mathrm{R}_{i}=\mathrm{K}_{ai}$ is satisfied.

\subsection{Holstein-Primakoff bosonization}\label{sec:HP}
In the HP approach we express the spin operators in terms of bosonic operators, which ensure that the commutators between spin operators are conserved. In the local reference frame, the ground state spin direction is always along $z$ and we thus apply the {\it local} Holstein-Primakoff transformation \cite{HP1940, Toth2015},
\begin{align}
    S^{\dprime +}_{ai} &= \sqrt{2S_a} \sqrt{1-\frac{a_{ai}^\dagger a_{ai}}{2S_a}} a_{ai}  \label{eq:hpS+},\\
    S^{\dprime -}_{ai} &= \sqrt{2S_a} a^\dagger_{ai} \sqrt{1-\frac{a_{ai}^\dagger a_{ai}}{2S_a}\label{eq:hpS-}},\\
    S^{\dprime z}_{ai} &= S_a - a^\dagger_{ai} a_{ai} \label{eq:hpSz},
\end{align}
where $S_a$ is the maximum eigenvalue of the spin operator at sublattice $a$. Applying this to Eq. \eqref{eq:Sprime-uv}, and truncating the square roots at third order in bosonic operators we obtain
\begin{align}
    \mathbf{S}^\prime_{ai}  &\approx \mathbf{u}^{*T}_a \big(\sqrt{\frac{S_a}{2}}a_{ai} - \frac{a^\dagger_{ai} a_{ai} a_{ai}}{4\sqrt{2S_a}} \big) \notag\\& + \mathbf{u}^T_a \big(\sqrt{\frac{S_a}{2}}a^\dagger_{ai} - \frac{a^\dagger_{ai} a^\dagger_{ai} a_{ai}}{4\sqrt{2S_a}} \big)   \notag\\& + \mathbf{v}^T_a(S_a - a^\dagger_{ai} a_{ai} ) \label{eq:Sprime-uv-hp}.
\end{align}
Using this, the Hamiltonian \eqref{eq:rot-heisenberg-model_sia} can be expanded up to fourth order in bosonic operators:
\begin{equation}
    \mathcal{H}_\mathrm{K} \approx \mathcal{E}^0_\mathrm{K}  + \mathcal{H}_\mathrm{K}^2 + \mathcal{H}_\mathrm{K}^4,
\end{equation}
where
\begin{align}
    \mathcal{E}^0_\mathrm{K} &= -N\sum_{a} (\mathbf{v}^T_a\mathrm{K}_a \mathbf{v}_a S_a(S_a+1),\\
    \mathcal{H}_\mathrm{K}^2 &= \sum_{ai} \Big(S_a\mathbf{v}^T_a \mathrm{K}_{a} \mathbf{v}_a (a^\dagger_{ai} a_{ai} + a_{ai} a^\dagger_{ai} )\notag\\& -\frac{S_a}{2} \big(  \mathbf{u}^{*T}_a \mathrm{K}_{a} \mathbf{u}^*_a a_{ai}a_{ai} + 
     \mathbf{u}^{*T}_a \mathrm{K}_{a} \mathbf{u}_a a_{ai}a^\dagger_{ai} \notag\\ &\qquad+
     \mathbf{u}^T_a \mathrm{K}_{a} \mathbf{u}^*_a a^\dagger_{ai}a_{ai} +
    \mathbf{u}^T_a \mathrm{K}_{a} \mathbf{u}_a a^\dagger_{ai} a^\dagger_{ai} \big)\Big),\label{eq:Hhp_2}\\
    \mathcal{H}_\mathrm{K}^4 =& \sum_{ai} \Bigl(-\mathbf{v}_a^T \mathrm{K}_{a} \mathbf{v}_a a^\dagger_{ai}a_{ai}a^\dagger_{ai}a_{ai} \notag\\&+\frac{1}{8}\Bigl( \mathbf{u}^{*T}_a \mathrm{K}_{a} \mathbf{u}^*_a  (2a^\dagger_{ai}a_{ai}a_{ai}a_{ai} + a_{ai} a_{ai}) \notag \\&\quad 
    +\mathbf{u}^{*T}_a \mathrm{K}_{a} \mathbf{u}_a (2a^\dagger_{ai} a^\dagger_{ai}a_{ai}a_{ai} + 4a^\dagger_{ai}a_{ai}) \notag \\
    &\quad+
    \mathbf{u}_a^T \mathrm{K}_{a} \mathbf{u}^*_b  2a^\dagger_{ai}a^\dagger_{ai}a_{ai}a_{ai}  \notag \\& \quad+ \mathbf{u}_a^T \mathrm{K}_{a} \mathbf{u}_a (2a^\dagger_{ai}a^\dagger_{ai}a^\dagger_{ai}a_{ai} + a^\dagger_{ai} a^\dagger_{ai}\Bigr).\label{eq:Hhp_4}
\end{align}
We note that normal ordering of the quartic terms generates more quadratic terms that were not taken into account in $\mathcal{H}_\mathrm{K}^2$ and will affect the zero-temperature dispersion. The bosonic operators are then Fourier transformed:
\begin{align}
    a_{a,\mathbf{q}} = \frac{1}{\sqrt{N}} \sum_i a_{ai} e^{i\mathbf{q}\cdot \mathbf{r}_i}, \quad a\in [1,N_a], \label{eq:a_q}
\end{align}
which results in fourth order terms like $a^\dagger_{a,\mathbf{q}}a^\dagger_{a,\mathbf{q^\prime}}a_{a,\mathbf{q^\dprime}}a_{a,\mathbf{q+q^\prime-q^\dprime}}$. These are subjected to a mean field decoupling leading to terms of the form $ a^\dagger_{a,\mathbf{q}}a_{a,\mathbf{q}}\langle a^\dagger_{a,\mathbf{q^\prime}}a_{a,\mathbf{q^\prime}} \rangle\delta_\mathbf{q,q^\dprime}$, where $\langle\ldots\rangle$ denotes the thermal average, which is assumed to vanish unless the bosonic operators contain the same wavevectors. Introducing the vector
\begin{align}
    \mathbf{a}_\mathbf{q} = 
    \begin{bmatrix}
        a_{a,\mathbf{q}} & a^\dagger_{b,\mathbf{-q}}
    \end{bmatrix}^T, \quad a,b \in [1,N_a],
\end{align}
Eq. \eqref{eq:heisenberg-model} can be expressed as
\begin{align}\label{eq:heisenberg-hp}
    \mathcal{H} &= \mathcal{E}_0 + \mathcal{E}^0_\mathrm{K} +\sum_{\mathbf{q}} \mathbf{a}^\dagger_{\mathbf{q}}  \big(\mathrm{H}^\mathrm{HP}_\mathbf{q} + \mathrm{H^{HP}_K}\big)\mathbf{a}_{\mathbf{q}}.
\end{align}
where $\mathcal{E}_0$ is a constant  and $\mathrm{H}^\mathrm{HP}_\mathbf{q}$ is the exchange contribution defined in appendix \ref{app:exchange_HP}. The single-ion anisotropy Hamiltonian, which is independent of $\mathbf{q}$, is given by
\begin{align}
    \mathrm{H}^\mathrm{HP}_\mathrm{K} &= 
    \frac{\delta_{ab}}{2}\begin{bmatrix}
        -\mathrm{A}^\mathrm{HP}_{a} +  \mathrm{C}^\mathrm{HP}_{a} & -\mathrm{B}^\mathrm{HP}_{a} \\
         -(\mathrm{B}^\mathrm{HP}_{a})^\dagger & -\mathrm{A}^\mathrm{HP}_{a} + \mathrm{C}^\mathrm{HP}_{a} 
    \end{bmatrix} \label{eq:H-hpK}
\end{align}
with
\begin{align}
    \mathrm{A}^\mathrm{HP}_{a} &= S_a\mathbf{u}_a^T\mathrm{K}_{a} \mathbf{u}^*_a  \Big(1 - \frac{1}{2S_a}(1 + 4\Phi_a)\Big),
    \\ 
    \mathrm{B}^\mathrm{HP}_{a} &= S_a\mathbf{u}_a^T\mathrm{K}_{a} \mathbf{u}_a \Big(1 - \frac{1}{4S_a}(1 +6\Phi_a)\Big), \\
    \mathrm{C}^\mathrm{HP}_{a} &=  S_a\mathbf{v}_a^T\mathrm{K}_{a} \mathbf{v}_a \Big(2 - \frac{1}{S_a}(1 + 4\Phi_a )\Big),
\end{align}
and the magnon occupation $\Phi_a$ is defined in Eq. \eqref{eq:PhiHP}. The terms in brackets $\frac{1}{S_a}(\dots)$ arise from the quartic Hamiltonian \eqref{eq:Hhp_4}. 
The isotropic invariance is preserved by these terms, which is easily seen by noting that $\mathbf{u}^{T}_a\mathbf{u}^*_a=2$, $\mathbf{u}^{T}_a\mathbf{u}_a=0$, and $\mathbf{v}^{T}_a\mathbf{v}_a=1$.
The magnon energies are obtained by (para)diagonalizing Eq. \eqref{eq:heisenberg-hp} \cite{Toth2015,Helimag3d2025}. The sublattice magnetization is then calculated as 
\begin{equation}
    \langle S^{\dprime z}_a \rangle = S_a - \Phi_a, \label{eq:hp-magnetization}
\end{equation}
where $\Phi_a$ must be calculated self-consistently from the magnon energies. The ground state magnetization, $S^{(0)}_a = \langle S^{\dprime z}_a \rangle_{T=0}$ can be obtained by evaluating Eq. \eqref{eq:hp-magnetization} at $T=0$ and satisfies $S^{(0)}_a \leq S_a$ where the equality only applies to ferromagnets. In principle, one should thus be able to calculate the critical temperature as the point where the sublattice magnetization vanishes. However, the method breaks down at elevated temperatures as the magnon energies become negative before the magnetization vanishes. The critical temperature may then be evaluated as the point where the derivative of magnetization with respect to temperature approaches $-\infty$. However, this behavior signifies a pathological feature  of the HP method, which may render results at high temperatures inaccurate.

In order to analyze the dependence of the magnon gap (at $T=0$), we consider the magnon dispersion at $T=0$. For a Bravais lattice one obtains
\begin{align}
    \omega^\mathrm{HP}_\mathbf{q} &= S\sqrt{{X^\mathrm{HP}_\mathbf{q}}^2 - |Y^\mathrm{HP}_\mathbf{q}|^2},\label{eq:whp_q}
\end{align}
with
\begin{align}
    X^\mathrm{HP}_\mathbf{q} &= \frac{\mathbf{u}^T\mathrm{J}^\prime_\mathbf{q}\mathbf{u}^*}{2} - \mathbf{v}^T\mathrm{J}^\prime_\mathbf{q} \mathbf{v} + \mathbf{u}^T\mathrm{K}\mathbf{u}^*\Big(1 - \frac{1}{S}(1+2\Phi^0)\Big)  \notag\\
    &- \mathbf{v}^T\mathrm{K}\mathbf{v} \Big( 2 - \frac{1}{S}(1+4\Phi^0)\Big) \label{eq.xhp_q}\\
    Y^\mathrm{HP}_\mathbf{q} &= \frac{\mathbf{u}^T\mathrm{J}^\prime_\mathbf{q}\mathbf{u}}{2} + \mathbf{u}^T\mathrm{K}\mathbf{u}\Big(1 -\frac{1}{4S}(1+6\Phi^0)\Big)\label{eq:yhp_q}.
\end{align}
$\Phi^0$ is the saturation magnon number, obtained by evaluating Eq. \eqref{eq:PhiHP} at $T=0$ 
and $\mathrm{J}^\prime_\mathbf{q}$ is defined in Eq. \eqref{eq:Jprime_q}.
Note that all the site indices have been dropped as we are considering a simple Bravais lattice. The magnon gap is defined as 
\begin{equation}
    \Delta = \omega^\textrm{HP}_{\mathbf{q}=0}\label{eq:delta-hp}
\end{equation}
and it is is straightforward to show that the $\frac{1}{S}$ terms yield zero temperature corrections to the magnon gap. For ferromagnets $\Phi^0$ vanishes, and the magnon gap in uniaxial systems can be expressed as
\begin{equation}
    \Delta = K(2S-1)
\end{equation}
where $K=\mathrm{K}^{zz}-\mathrm{K}^{xx}$.
This is obtained by evaluating Eq. \eqref{eq:delta-hp} with $\mathbf{u}=[1,i,0]$ and $\mathbf{v}=[0,0,1]$ (ground state magnetization along the $z$-axis).
We see that the magnon gap vanishes in the case of $S=1/2$, which is the expected result. However, at finite temperatures this no longer holds and for tri-axial systems the magnon dispersion will acquire a dependence on the single-ion anisotropy at $T=0$ for $S=1/2$ and may even lead to instabilities (negative magnon energies). 
The basic reason for this failure is the truncation of $S^{\dprime \pm}_{ai}$ in Eqs. \eqref{eq:hpS+}-\eqref{eq:hpS-}, which becomes inaccurate if the anisotropy tensor couples spins that are not along the local $z$-axis.

\subsection{Green's function method}\label{sec:Gfn}
The RPA Green’s function approach has previously been shown to yield rather accurate predictions for critical temperatures in single-$Q$ bulk systems \cite{Helimag3d2025}. Here, we extend the RPA to include single-ion anisotropy. The method is then shown to fail rather strikingly in the case of $S=1/2$ and is expected to overestimate spinwave gaps in general. We then proceed with the Callen decoupling applied to single-ion anisotropies and show that the failure of RPA is largely remedied in this approach. 

In general, the magnon energies appear as the poles of the dynamic susceptibility. Here we consider two of the (locally) transverse components, which we will just refer to as retarded Green's functions. They are defined by:
\begin{align}
    G^{+-}_{abij}(t) &= -i\theta(t) \langle [S^{\dprime +}_{ai}(t), S^{\dprime -}_{bj}]\rangle \equiv \langle\langle  S^{\dprime +}_{ai}(t) ;S^{\dprime -}_{bj} \rangle \rangle \label{eq:G_+-},\\
    G^{--}_{abij}(t)  &= -i\theta(t)\langle[ S^{\dprime -}_{ai}(t), S^{\dprime -}_{bj}] \rangle \equiv \langle\langle S^{\dprime -}_{ai}(t) ;S^{\dprime -}_{bj} \rangle \rangle, \label{eq:G_--}
\end{align}
where $\theta(t)$ is the Heaviside step-function. The temporal Fourier transforms are denoted by
\begin{align}
    G_{abij}^{\pm-}(\omega) &= \int_{-\infty}^\infty dt\, e^{i(\omega+i\epsilon)t}\langle\langle S^{\dprime \pm}_{ai}(t) ; S^{\dprime -}_{bj} \rangle \rangle  \notag \\&\equiv  \langle\langle S^{\dprime \pm}_{ai} ; S^{\dprime -}_{bj} \rangle \rangle_\omega, \label{eq:G(w)}
\end{align}
where $\epsilon$ is an infinitesimal positive frequency that ensures convergence of the integral. Evaluating the equation of motion for the Green's functions will generate higher order (more than two operators) Green's functions that need to be approximated. In particular, the contributions from single-ion anisotropy appear in higher order Green's functions of the form 
\begin{equation}
\langle\langle [S^{\dprime \pm}_{ai},\mathcal{H}_\mathrm{K}]; S^{\dprime -}_{bj}\rangle\rangle_\omega.
\end{equation}
In the following we consider two decoupling approaches: the Random Phase Approximation (RPA) and the Callen decoupling (CD) scheme.


\subsubsection{Random Phase Approximation}
In the RPA individual spin fluctuations along the local $z$-axis $S^{\dprime z}_{ai}$ are assumed to be small and the operator is replaced by its ensemble average $\langle S^{\dprime z}_a \rangle$. This means that the three-operator Green's functions are approximated by
\begin{align}\label{rpa-approx}
    \langle \langle S^{\dprime \pm}_{ai} S^{\dprime z}_{ai}; S^{\dprime -}_{bj} \rangle \rangle_\omega &\approx \big\langle S^{\dprime z}_a \rangle    G^{\pm-}_{abij}(\omega).
\end{align}
Applying this, the equations of motion for the Green's functions can be written as a generalized eigenvalue equation
\begin{align}
    \big((\omega + i\epsilon)\mathrm{I} - \big(\mathrm{H}^\mathrm{RPA}_\mathbf{q} + \mathrm{H}^\mathrm{RPA}_\mathrm{K} \big) \big) \mathrm{G}_\mathbf{q}(\omega) &= \zeta \label{eq:eqnofmotion},
\end{align}
where 
\begin{align}
    \mathrm{G}_\mathbf{q}(\omega) &=
    \sum_i 
    \begin{bmatrix}
         G_{ab0i}^{+-}(\omega) & G_{ab0i}^{--}(\omega) 
    \end{bmatrix}^T e^{i\mathbf{q}\cdot\mathbf{r}_i},\\
    \zeta &=
    \begin{bmatrix}
         2\langle S^{\dprime z}_a \rangle \delta_{ab}  & 0  
    \end{bmatrix}^T, \quad a,b \in [1,N_a].
\end{align}
Here $\mathrm{H}^\mathrm{RPA}_\mathbf{q}$ contains the contribution from the exchange tensor, which has been derived previously \cite{Helimag3d2025} and is stated in appendix \ref{app:RPA}. The single-ion anisotropy contribution, $\mathrm{H}^\mathrm{RPA}_\mathrm{K}$ is given by 
\begin{align}
    \mathrm{H}^\mathrm{RPA}_\mathrm{K} &=
    \delta_{ab}
    \begin{bmatrix}
        -\mathrm{A}^\mathrm{RPA}_{a} +  \mathrm{C}_{a}^\mathrm{RPA} & -\mathrm{B}_{a}^\mathrm{RPA} \\
         (\mathrm{B}^\mathrm{RPA}_{a})^\dagger & \mathrm{A}^\mathrm{RPA}_{a} - \mathrm{C}^\mathrm{RPA}_{a} 
    \end{bmatrix}\label{eq:Hrpa-sia}
\end{align}
with
\begin{align}
    \mathrm{A}^\mathrm{RPA}_{a} &= \langle S^{\dprime z}_a \rangle \mathbf{u}^{T}_a \mathrm{K}_{a}\mathbf{u}^*_a \label{eq:Ak-rpa},
    \\
    \mathrm{B}^\mathrm{RPA}_{a} &= \langle S^{\dprime z}_a \rangle\mathbf{u}^{T}_a \mathrm{K}_{a}\mathbf{u}_a,
    \\
    \mathrm{C}^\mathrm{RPA}_{a} &= 2 \langle S^{\dprime z}_a \rangle \mathbf{v}^{T}_a \mathrm{K}_{a}\mathbf{v}_a \label{eq:Ck-rpa}.
\end{align}
It is straightforward to verify that isotropic invariance is conserved by $\mathrm{H}^\mathrm{RPA}_\mathrm{K}$.

Similar to HP, we can calculate the magnon energies at $T=0$ for a Bravais lattice using RPA. The result is
\begin{align}
    \omega^\mathrm{RPA}_\mathbf{q} &= S^{(0)}\sqrt{{X^\mathrm{RPA}_\mathbf{q}}^2 - |Y^\mathrm{RPA}_\mathbf{q}|^2},
\end{align}
where $S^{(0)}$ is the saturation magnetization calculated as in Eq. \eqref{eq:Sz0} and 
\begin{align}
    X^\mathrm{RPA}_\mathbf{q} &= \frac{\mathbf{u}^T(\mathrm{J}^\prime_\mathbf{q} + 2\mathrm{K})\mathbf{u}^*}{2} - \mathbf{v}^T(\mathrm{J}^\prime_\mathbf{q} + 2\mathrm{K})\mathbf{v}, \\
    Y^\mathrm{RPA}_\mathbf{q} &= \frac{\mathbf{u}^T(\mathrm{J}^\prime_\mathbf{q} + 2\mathrm{K})\mathbf{u}}{2},
\end{align}
where $\mathrm{J}^\prime_\mathbf{q}$ is defined in Eq. \eqref{eq:Jprime_q}.
It can be observed that RPA predicts a finite gap for $S=1/2$ and fails to capture the irrelevance of singe-ion anisotropy in this case. This is a serious flaw of the method, which we address by turning to the CD scheme for the single-ion anisotropy.

\subsubsection{Callen decoupling}
As RPA assumes the fluctuations in $S^{\dprime z}_{ai}$ to be small, the large fluctuations that would become relevant at higher temperatures are disregarded {\it a priori}. The CD attempts to generalize the RPA mean-field approach in such a way that fluctuations in $S^{\dprime z}_{ai}$ become important at high temperatures \cite{Callen1963,Callen1964-afm}. However, comparing results with quantum Monte Carlo simulations for $S=1/2$ ferromagnets reveals that the method actually yields worse results than RPA \cite{Frobich2006}. For higher spins it is expected to become more accurate and for itinerant systems (treated in the classical limit) the CD was found to be more accurate than RPA \cite{BeyondRPA2025}. Moreover, it has been shown that the CD predicts much better results than RPA when it is solely applied to single-ion anisotropy in ferromagnets \cite{BeyondRPA2025,Frobich2006}. One reason for this is that it  predicts the correct lack of order in $S=1/2$ and is not subject to the overestimation of spinwave gaps inherent to RPA. As the failure of RPA for single-ion anisotropy is not limited to ferromagnets, we derive a single-$Q$ generalization of the CD for single-ion anisotropy. Combined with the RPA Hamiltonian for the exchange thus yields the hybrid RPA+CD approach. 

The central idea in CD is parameterizing the three different representations of the $S_{ai}^{\dprime z}$ operator:
\begin{align}
    S_{ai}^{\dprime z} &= S_a(S_a+1) - S_{ai}^{\dprime -}S_{ai}^{\dprime +} - {S_{ai}^{\dprime z}}^2 \label{eq:Sz-cd1},\\
    S_{ai}^{\dprime z} &= -S_a(S_a+1) + S_{ai}^{\dprime +}S_{ai}^{\dprime -} + {S_{ai}^{\dprime z}}^2 \label{eq:Sz-cd2},\\
    S_{ai}^{\dprime z} &= \frac{S_{ai}^{\dprime +}S_{ai}^{\dprime -} - S_{ai}^{\dprime -}S_{ai}^{\dprime +}}{2} \label{eq:Sz-cd3}.
\end{align}
Introducing the parameter $\alpha\in\{-1,0,1\}$ these equations can be combined into a single one:
\begin{align}\label{eq:Sz-cd}
    S_{ai}^{\dprime z} &= \alpha (S_a(S_a+1)- {S_{ai}^{\dprime z}}^2) + \frac{1}{2}(1-\alpha)S_{ai}^{\dprime +}S_{ai}^{\dprime -}\notag\\& - \frac{1}{2}(1+\alpha)S_{ai}^{\dprime -}S_{ai}^{\dprime +}.
\end{align}
The point is now that the decoupling will depend on the choice of $\alpha$ and Callen argued that $\alpha$ should be chosen differently for the low and high temperature regimes. Applying the parameterized representation of $S_{ai}^{\dprime z}$, the higher-order Green's functions can be symmetrically decoupled as
\begin{align}\label{eq:cd-approx}
    \langle \langle S^{\dprime \pm}_{ai} S^{\dprime z}_{ai}; S^{\dprime -}_{bj} \rangle \rangle_\omega &\approx \big(\langle S^{\dprime z}_a \rangle    - \alpha \langle S^{\dprime \pm}_{ai} S^{\dprime \mp}_{ai} \rangle\big) G^{\pm-}_{abij}(\omega),
\end{align}
and the parameter $\alpha$ can be chosen as $\frac{\langle S_a^{\dprime z} \rangle}{2S_a^2}$ satisfying the physical requirements discussed in Ref. \cite{Callen1963}.
Applying this to Eq. \eqref{eq:cd-approx}, the CD single-ion anisotropy Hamiltonian is obtained as 
\begin{equation}
    \big[\mathrm{H}^\mathrm{CD}_\mathrm{K}\big]_{\alpha\beta} = \big[\mathrm{H}^\mathrm{RPA}_\mathrm{K}\big]_{\alpha\beta} \Big(1- \frac{\langle S^{\dprime z}_a \rangle}{2S_a^2}(1+2\Phi_a)\Big)\delta_{\alpha\beta}, \label{eq:Hcd-k}
\end{equation}
where $\alpha,\beta\in[1,2N_a]$, $\mathrm{H}^\mathrm{RPA}_\mathrm{K}$ is defined in Eq. \eqref{eq:Hrpa-sia} and $\Phi_a$ is the magnon occupation number defined in Eq. \eqref{eq:Phi_a}. 
Similar to RPA,
the magnon energies within the RPA+CD method can be obtained by solving the eigenvalue equation 
\begin{equation}
     \big((\omega + i\epsilon)\mathrm{I} - \big(\mathrm{H}^\mathrm{RPA}_\mathbf{q} + \mathrm{H}^\mathrm{CD}_\mathrm{K} \big) \big) \mathrm{G}_\mathbf{q}(\omega) = \zeta ,
\end{equation}
with $\mathrm{H}^\mathrm{RPA}_\mathbf{q}$ being the RPA exchange contribution in Eq. \eqref{eq:H_rpa} and $\mathrm{H}^\mathrm{CD}_\mathbf{q}$ is given by Eq. \eqref{eq:Hcd-k}.

Analyzing the behavior of single-ion anisotropy in RPA+CD, we find
\begin{equation}\label{eq:H0cd}
    \mathrm{H}^\mathrm{CD}_\mathrm{K}(T=0) \propto \Big(1- \frac{(S_a-\Phi^0_a)(1+2\Phi^0_a)}{2S_a^2}\Big)
\end{equation}
where $\Phi^0$ is defined in Eq. \eqref{eq:Phi0}. It is observed that for $S=1/2$, this contribution is close to being independent of $\mathrm{K}_a$, as the quantity in bracket is close to (exactly for ferromagnets) zero. Importantly, it is RPA+CD that yields the closest agreement to the true behavior (irrelevance) of single-ion anisotropy in $S=1/2$ systems when comparing the three methods considered here. 



\subsubsection{Magnetization and Critical temperature}
In the Green's function approach, the (sublattice) magnetization is given by \cite{Callen1963,Helimag3d2025,BeyondRPA2025}
\begin{equation}\label{eq:magnetization}
    \langle S_a^{\dprime z}  \rangle = \frac{(S_a-\Phi_a)(1+\Phi_a)^{2S_a+1} + (S_a+1+\Phi_a)\Phi_a^{2S_a+1}}{(1+\Phi_a)^{2S_a+1} - \Phi_a^{2S_a+1}},
\end{equation}
where $\Phi_a$ is the site-resolved magnon occupation calculated from the fluctuation-dissipation theorem as
\begin{equation}\label{eq:Phi_a}
    \Phi_a = \frac{1}{N}\sum_{\eta,\mathbf{q}} U_{a \eta \mathbf{q}} n_\mathrm{B}(\omega_{\eta,\mathbf{q}}) U^{-1}_{\eta a \mathbf{q}},
\end{equation}
where and $U_{a\eta\mathbf{q}}$ is the matrix that diagonalizes 
\begin{equation}
    \mathrm{H}_\mathbf{q}^\mathrm{RPA} + \mathrm{H}_\mathrm{K}^\mathrm{RPA/CD}
\end{equation}
and $\omega_{\eta, \mathbf{q}}$ are the eigenvalues. We note that in Eqs. \eqref{eq:magnetization}-\eqref{eq:Phi_a}, $a\in\{1,\ldots,N_a\}$, but the sum in \eqref{eq:Phi_a} runs over all ($2N_a$) eigenfrequencies for a given $\mathbf{q}$. The bose factor $n_\mathrm{B}(\omega_{\eta,\mathbf{q}})$ is given by
\begin{equation}\label{eq:bosefactor}
    n_\mathrm{B}(\omega_{\eta,\mathbf{q}}) = \frac{1}{e^{\omega_{\eta\mathbf{q}}/k_\mathrm{B}T}-1},
\end{equation}
and is thus evaluated at positive as well as negative eigenfrequencies although only the positive eigenvalues are interpreted as magnon energies. For each temperature, the magnetization must be calculated self-consistently since the Hamiltonian depends on the Bose factors and the magnon energies are thus renormalized (softened) at finite temperatures.

In the limit of small $\Phi$, Eq. \eqref{eq:magnetization} can be expanded as
\begin{equation}
    \langle S^{\dprime z}_a \rangle = S_a - \Phi_a + \mathcal{O}(\Phi^{2S_a+1}).
\end{equation}
At zero temperature, this reduces to the saturation magnetization, $S^{(0)}_a$, given by
\begin{align}\label{eq:Sz0}
    S^{(0)}_a = \langle S^{\dprime z}_a \rangle_{T=0} 
    = S_a - \Phi^0_a
\end{align}
with
\begin{align}
    \Phi^0_a &=  -\frac{1}{N}\sum_{\mathbf{q}} \sum_{n=1}^{N_a} U_{a n \mathbf{q}} U^{-1}_{n a \mathbf{q}} , \label{eq:Phi0}
\end{align}
where we used 
\begin{align}
    n^{T=0}_\mathrm{B}(\omega_{\eta\mathbf{q}}) &= 
    \begin{cases}
        -1, \quad& \eta \in [1,N_a] \\
        0, \quad &\eta \in (N_a,2N_a]
    \end{cases}.
\end{align}
Note that the above equation assumes that all the eigenvectors are sorted according to the ascending eigenvalues. Except for ferromagnets, the saturation magnon number will be non-vanishing and $S^{(0)}_a<S_a$. In contrast to the HP approach, the magnetization calculated from Eq. \eqref{eq:magnetization} will yield a function of temperature, which smoothly approaches zero as we reach the critical temperature. For systems with equivalent site magnetization, $\langle S^{\dprime z}_a \rangle = \langle S^{\dprime z} \rangle$ $\forall a$, we can obtain an analytical expression for the RPA critical temperature by expanding Eq. \eqref{eq:magnetization} in the limit of $\langle S^{\dprime z} \rangle \rightarrow 0$:
\begin{align}
    k_\mathrm{B}T^\mathrm{RPA}_\mathrm{C} &= \frac{S(S+1)}{3}\Bigg( \frac{\langle S^{\dprime z} \rangle}{N}\sum_{\eta \mathbf{q}} \frac{U_{0 \eta \mathbf{q}} U^{-1}_{\eta 0 \mathbf{q}}}{\omega_{\eta,\mathbf{q}}} \Bigg)^{-1} \label{eq:Tc-gen}.
\end{align}
Since $\omega_{\eta,\mathbf{q}}$ is proportional to $\langle S^{\dprime z} \rangle$, the expression effectively no longer depends on it.
This is a generalization of the critical temperature of a Bravais lattice derived in our previous work \cite{Helimag3d2025} to non-Bravais lattices with equivalent spin sites. For the RPA+CD approach, the critical temperature must be obtained from a numerical (self-consistent) calculation of the magnetization curve.

\subsection{Critical temperatures in the limit of small anisotropies}\label{sec:limitingcases}
The Mermin-Wagner theorem has established that  magnetic anisotropy plays a pivotal role in the magnetic order in 2D materials. 
Here we study the behavior of the critical temperature in the limit of small magnon gaps, corresponding to the regime of weak magnetic anisotropy. We analyze the cases of simple ferromagnetic and anti-ferromagnetic order with uniaxial anisotropy using RPA.

For ferromagnetic order, the magnon dispersion is quadratic in the vicinity of $\mathbf{q}=0$ and the magnon gap ($\Delta_\textrm{FM}$) is linear with respect to the anisotropy strength. Assuming an isotropic lattice, it can be expressed as
\begin{equation}
    \omega^\mathrm{FM}_\mathbf{q}=\langle S^{\dprime z} \rangle(bq^2+ \Delta_\mathrm{FM}).
\end{equation}
For $\omega_{\mathbf{q}}\ll k_\mathrm{B}T$, the Bose factor in Eq. \eqref{eq:bosefactor} can be expanded as
\begin{equation}
    n_\mathrm{B}(\omega_{\eta,\mathbf{q}}) \approx \begin{cases}
        -\frac{k_\mathrm{B}T}{\omega_\mathbf{q}} \quad \eta=1 \\
        +\frac{k_\mathrm{B}T}{\omega_\mathbf{q}} \quad \eta=2
    \end{cases}.
\end{equation}
If the magnon gap becomes small, the largest contribution of the magnon number Eq. \eqref{eq:Phi_a} will originate from the vicinity of $\mathbf{q}=0$. This contribution can be separated from the rest such that
\begin{align}
    \Phi = \frac{2}{q_c^2\langle S^{\dprime z} \rangle}\int_0^{q_c} k_\mathrm{B}T\frac{qdq}{bq^2+\Delta}_\mathrm{FM}+\sum_{|\mathbf{q}| > q_c}\ldots,
\end{align}
where $q_c$ is some small cutoff and the second term will become irrelevant if the anisotropy is sufficiently small. The magnetization in Eq. \eqref{eq:magnetization} can be expanded in the limit of large $\Phi$ ($\langle S^{\dprime z} \rangle \rightarrow 0$) as
\begin{align}
    \langle S^{\dprime z} \rangle &\approx \frac{S(S+1)}{3\Phi}, \label{eq:Sz-nearTc}
    \\&= \frac{S(S+1)}{3} \Big(  \frac{2}{q_c^2\langle S^{\dprime z} \rangle}\int_0^{q_c} k_\mathrm{B}T\frac{qdq}{bq^2+\Delta_\mathrm{FM}} \Big)^{-1},
\end{align}
which becomes exact as $T\rightarrow T_\mathrm{C}$ and one obtains
\begin{align}
     k_\mathrm{B}T^\mathrm{RPA}_\mathrm{C} &\propto \Big[\mathrm{ln}\Big(\frac{ bq_c^2+\Delta_\mathrm{FM}}{\Delta_\mathrm{FM}}\Big)\Big]^{-1}.
\end{align}
The Curie temperature thus acquires a logarithmic dependence on the gap in the limit of small gaps.

For anti-ferromagnetic order, we consider the simple case of a quadratic lattice. In the vicinity of $\mathbf{q}=0$, the magnon dispersion becomes hyperbolic:
\begin{align}
    \omega^\mathrm{AFM}_\mathbf{q} = \langle S^{\dprime z} \rangle\sqrt{cq^2 + \Delta_\mathrm{AFM}^2}.
\end{align}
Performing the same steps as above in the limit of small anisotropy then yields
\begin{align}
     k_\mathrm{B}T^\mathrm{RPA}_\mathrm{N} &\propto\Big[\ln\Big(\frac{ cq_c^2+\Delta_\mathrm{AFM}^2}{\Delta_\mathrm{AFM}^2}\Big)\Big]^{-1}.
\end{align}
Note that for antiferomagnets  an additional term proportional to $(cq^2 + \Delta_\mathrm{AFM}^2)^{-1/2}$ arises from the eigenvectors $U_{a\eta\mathbf{q}}U^{-1}_{\eta a\mathbf{q}}$ when the magnon number \eqref{eq:Phi_a} is evaluated.
                                                                 
\begin{figure}
    \includegraphics[scale=0.3]{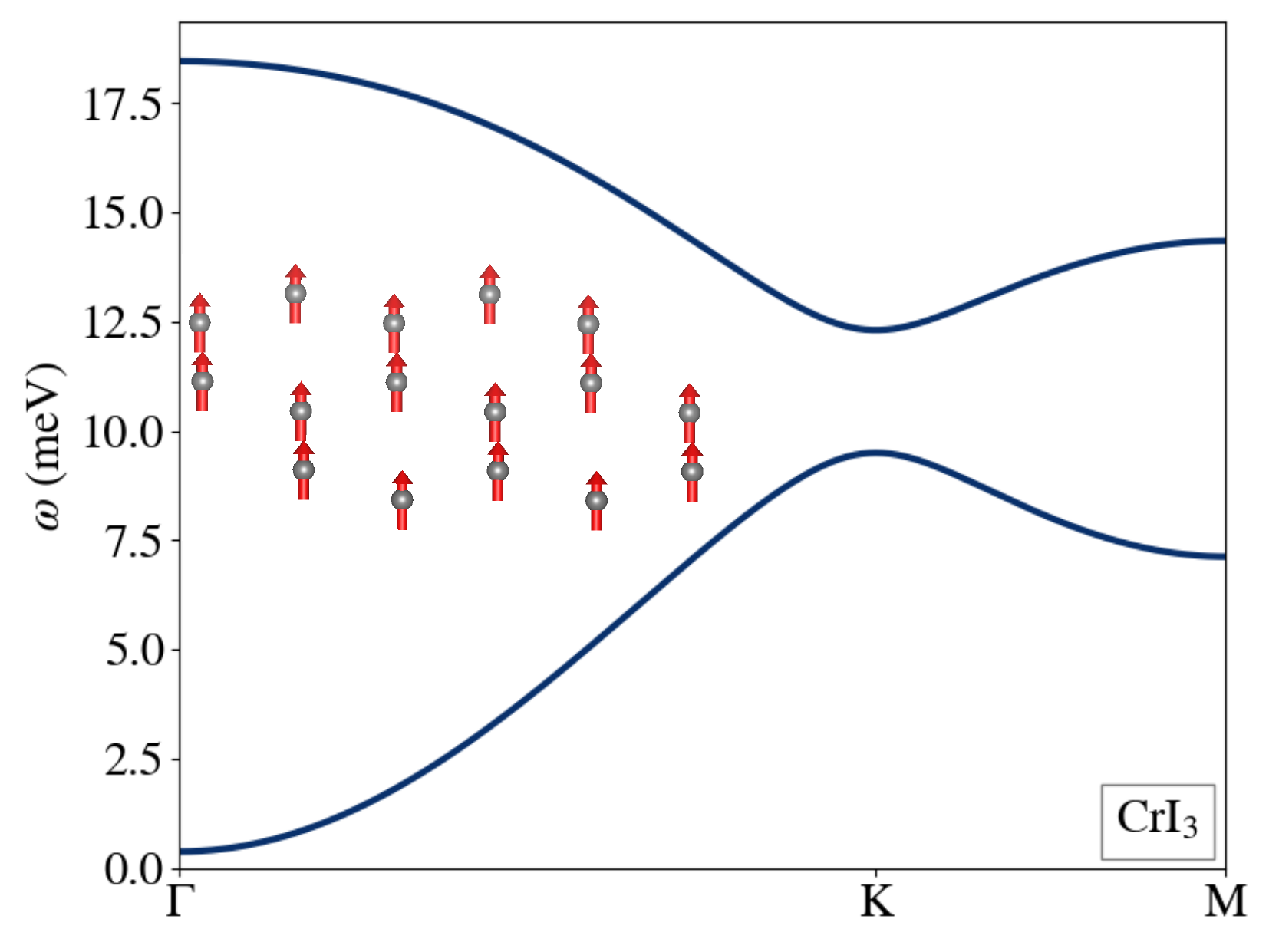}
    \caption{The magnon dispersion of monolayer CrI$_3$ calculated using bulk INS parameters from Ref. \cite{Chen-2021-cri3}. A DM interaction of 0.09 meV in the out-of-plane direction opens a topological gap at the high symmetry point $\textrm{K}$. The inset shows the magnetic lattice of CrI$_3$.}
    \label{fig:cri3}
\end{figure}

\subsection{Dipole interactions}\label{sec:dipole}
In addition to exchange and single-ion anisotropy, long range dipole–dipole interactions can also play an important role in two-dimensional magnetic materials. These interactions arise from the magnetic dipole moments of the ions and decay as $1/r^3$, leading to a long range contribution to the spin Hamiltonian. In the extended Heisenberg model, the dipolar term can be written as
\begin{equation}
    \mathcal{H}_\textrm{dip} = \frac{\mu_0}{8\pi}\sum_{abij} \Big( \frac{\mathbf{m}_{ai}\cdot\mathbf{m}_{bj}}{|\mathbf{r}_{abij}|^3} - \frac{3 (\mathbf{m}_{ai}\cdot\mathbf{r}_{abij})(\mathbf{m}_{bj}\cdot\mathbf{r}_{abij})}{|\mathbf{r}_{abij}|^5}\Big),
\end{equation}
where $\mu_0$ is the vacuum permeability. The magnetic moment is related to the spin by $\mathbf{m}_{ai}=-g\mu_\mathrm{B}\mathbf{S}_{ai}$, where $g$ is the spin $g$-factor and $\mu_\mathrm{B}$ is the Bohr magneton. This term can introduce long range anisotropic couplings and may play a role in stabilizing magnetic order in 2D in systems when spin-orbit coupling is weak. If the magnitude of the dipole-dipole interaction is comparable to the exchange energies, it may also modify the magnon spectra. In the methodology introduced above, the dipole interactions can be incorporated as a simple addition to the exchange tensor:
\begin{align}
    \mathrm{J}_{ab\mathbf{q}} &\rightarrow \mathrm{J}_{ab\mathbf{q}} - \mathrm{D}_{ab\mathbf{q}}, \\
     [\mathrm{D}_{ab\mathbf{q}}]^{\mu\nu} &= \frac{D}{N}\sum_{ij} \Big( \frac{\delta_{\mu\nu}}{|\mathbf{r}_{abij}|^3} - \frac{3[\mathbf{r}_{abij}]^\mu [\mathbf{r}_{abij}]^\nu}{|\mathbf{r} _{abij}|^5} \Big) e^{i\mathbf{q}\cdot \mathbf{r}_{ij}}\label{eq:DipoleTensor},\\
    D &= \frac{\mu_0}{4\pi}g^2\mu_\mathrm{B}^2. 
\end{align}
Since the dipolar interaction is inherently long ranged, the dipole tensor in Eq. \eqref{eq:DipoleTensor} must be evaluated to sufficiently large distances until the summation converges. Unlike the single-ion anisotropy, the dipolar anisotropy is dispersive ($q$-dependent) and therefore manifests itself differently in the magnon spectrum. The long range nature of the interaction also implies that the Mermin-Wagner theorem does not apply, and easy-plane order in 2D can be stabilized at finite temperatures by the dipole-dipole interaction even if it does not open a gap in the magnon spectrum \cite{bruno_spin-wave_1991}.

\section{Results}\label{sec:results}
We have introduced three different approaches to study the thermal renormalization of magnon energies and critical temperatures for single-$Q$ states including single-ion anisotropy. We now benchmark the methods on different 2D magnetic materials using exchange constants obtained from inelastic neutron scattering experiments on their bulk parent compounds. In particular, we take CrI$_3$, MPS$_3$ (M=Ni, Mn, Fe) and CrSBr, and show that the Heisenberg parameters reproduce the experimental magnon dispersions at zero temperature. We then calculate the critical temperature using HP, RPA, RPA+CD and classical Monte Carlo simulations and compare the results with experimental values.

\subsection{CrI$_3$}
CrI$_3$ was the first material shown to exhibit long-range ferromagnetic order in the monolayer limit \cite{Huang2017}. The Cr atoms form a honeycomb lattice and exhibits strong easy-axis uniaxial anisotropy. In fact, the three-fold rotational symmetry associated with each magnetic site renders the single-ion anisotropy tensor diagonal with $\mathrm{K}_a^{xx}=\mathrm{K}_a^{yy}$, and we are free to set these in-plane components to zero. The single-ion anisotropy is thus completely characterized by $\mathrm{K}_a^{zz}$. Since its discovery, there has been considerable interest in this material -
 theoretical as well as experimental 
\cite{Soriano2020,Chen2018a}. The Cr atoms are in the 3+ oxidation state and are well represented by $S=3/2$ sites in a Heisenberg model. Since CrI$_3$ is ferromagnetic, there is no reduction in the saturation magnetization, which is simply given by $S^{(0)}_a = 3/2$ as well. The Heisenberg exchange and anisotropy parameters extracted from inelastic neutron scattering (INS) experiments of bulk CrI$_3$ \cite{Chen-2021-cri3} are listed in Tab. \ref{table:1-J_n}. As we are interested in the monolayer limit, the interlayer exchange is not included in our calculations, but we assume that the Heisenberg parameters for a single monolayer are well approximated by those of the bulk material. The Dzyaloshinskii–Moriya (DM) interaction in CrI$_3$ is absent in the nearest neighbor bond (due to an inversion center), but appears between next nearest neighbors. It is reported to be relatively weak compared to the exchange constants, with a magnitude of $0.09$ meV \cite{Chen-2021-cri3}, but gives rise to a sizable gap between the acoustic and optical magnon branches. At the level of linear spin-wave theory it is only the out-plane component that opens such a gap and the remaining components are typically disregarded although they are not forbidden by symmetry in CrI$_3$. The gap between the magnon branches is topological (finite Chern number) and has been discussed extensively in the literature \cite{Chen-2021-cri3,Mook2021,Yaroslav2016}. In the present tensor notation, these out-of-plane DM component enter as intra-sublattice components $\mathrm{J}^{xy}_{aaij}/2=-\mathrm{J}^{yx}_{aaij}/2$, where $i$ and $j$ denote neighboring cells. Since single-ion anisotropy contributes differently to the magnon dispersion for HP and RPA treatments, the reported experimental values depend on which method the dispersion was fitted to. We have thus recalculated the uniaxial single-ion anisotropy constant, $\mathrm{K}^{zz}_a$, directly from the experimentally observed spin-wave gap, obtaining a value of 0.184 meV for RPA+CD (and HP) and 0.123 meV for RPA. This recalculation ensures a consistent treatment of anisotropy within our formalism and ensures that the calculated spinwave dispersion agrees with the experimental one regardless of the methodology. We note that a similar analysis was performed in our previous work on CrI$_3$ \cite{BeyondRPA2025}, but in that case we employed exchange constants extracted from optical magnon energies and the spin-wave gap of bilayer CrI$_3$ reported in Ref. \cite{Cenker2021}. In contrast, here we base our analysis on the full magnon spectrum of bulk CrI$_3$ measured via INS experiments, providing a more comprehensive and accurate parameter set.
We thus assume that the interlayer exchange constants in the monolayer are the same as those in bulk. Since the monolayer exchange constants are not experimentally known we cannot rigorously justify this assumption. However, the fact that the energy of the optical magnons in bilayer CrI$_3$  \cite{Cenker2021} coincides with that of bulk CrI$_3$  \cite{Chen-2021-cri3} implies that the weak interlayer van der Waals interactions do not influence the interlayer exchange constants.

The magnon dispersion along the high-symmetry path $\Gamma$–K–M (with INS parameters from Ref. \cite{Chen-2021-cri3}) is shown in Fig. \ref{fig:cri3} and the corresponding critical temperatures obtained using MC, HP, RPA, and RPA+CD methods are summarized in Tab. \ref{table:Tc}. It is observed that the RPA+CD approach yields the closest agreement to the experimental critical temperature, which is consistent with previous work \cite{BeyondRPA2025} based on parameters from bilayer CrI$_3$.
\begin{figure}
    \includegraphics[scale=0.45]{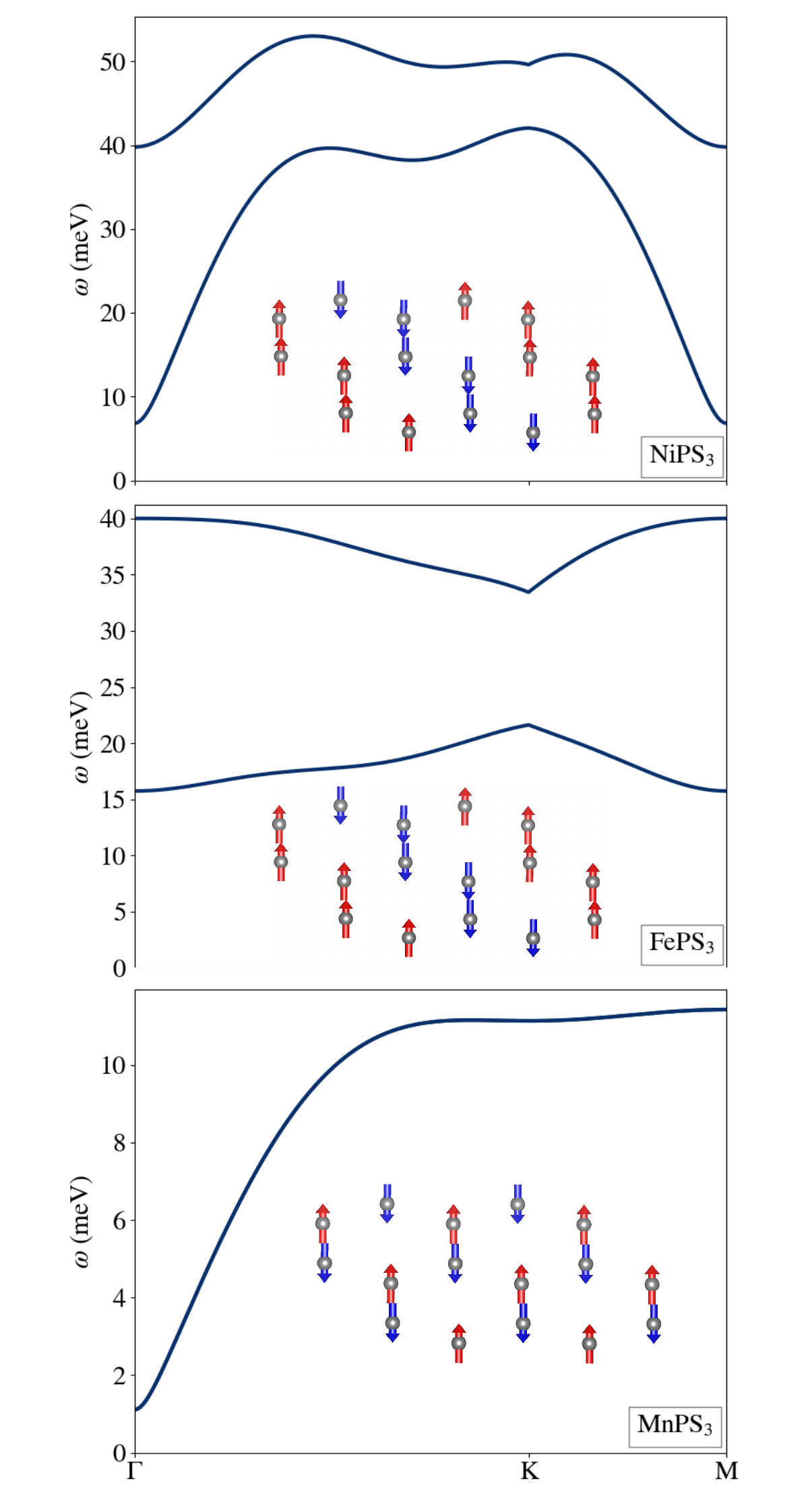}
    \caption{The magnon dispersion relations of monolayer MPS$_3$ (M=Ni, Fe, Mn) anti-ferromagnets with Heisenberg parameters taken from INS \cite{Wildes2018}. The inset shows the magnetic lattices with the corresponding ground state magnetic order.}
    \label{fig:MPS3}
\end{figure}

\subsection{MPS$_3$}
The MPS$_3$ (M=Ni, Mn, Fe) compounds comprise a family of layered transition-metal thiophosphates that crystallize with a honeycomb arrangement of transition-metal ions within each layer. Here, we consider NiPS$_3$, FePS$_3$ and MnPS$_3$, which gives rise to layered anti-ferromagnetic ordering with uniaxial anisotropy \cite{Vasudevan1992,Jernberg1984,Wildes1998,Wildes2015,Wildes2016}. NiPS$_3$ and FePS$_3$ exhibit a zig-zag magnetic ordering in the $ab$-plane with an in-plane ordering vector $\mathbf{Q}=[1/2,0]$, where spins in the unit cell are parallel with $S=1$ and $S=2$ respectively. MnPS$_3$, on the other hand exhibits Néel order, where all the nearest neighbors are anti-parallel and is represented by a $\mathbf{Q}=[0,0]$ in-plane ordering vector with anti-parallel spins in the primitive unit cell and $S=5/2$. The exchange and anisotropy parameters of these compounds have been studied extensively in the literature using both experimental \cite{Wildes1998,Wildes2016,Wildes2018,Wildes2015} and first principles approaches \cite{Olsen_2021, Kim2021-mps3}. Similar to the case of CrI$_3$, symmetry dictates that the single-ion anisotropy is completely characterized by $\mathrm{K}_a^{zz}=\mathrm{K}^{zz}$. For the purpose of benchmarking our methods, we use the Heisenberg parameters obtained from the bulk INS data reported in Ref. \cite{Wildes2018}, and the values are tabulated in Tab. \ref{table:1-J_n}. The calculated magnon dispersions for the three compounds are shown in fig. \ref{fig:MPS3}, along with the magnetic structure. Since these materials are anti-ferromagnets, the exchange constants ($J_n$) from the experimental fit performed with linear spin wave theory differ from RPA/RPA+CD by a factor of $S_a/\langle S^{\dprime z}_a\rangle$. To avoid confusion, we have retained the original unscaled exchange constants and assumed $S^{(0)}_a=S_a$ in our calculations.

\begin{table}[t]
    \centering
    \begin{tabular}{ |p{1.4cm}|p{1.2cm}|p{1.2cm}|p{1.2cm}|p{2.2cm}|}
        \hline
        & $J_1$  & $J_2$  & $J_3$ & $\mathrm{K}^{zz}$ \\
        \hline
        CrI$_3$ &2.11 &0.11 &-0.10 &0.123 (0.184)\\
        NiPS$_3$ &3.8 &-0.2 &-13.80 &0.3 (0.6)\\
        FePS$_3$ &2.92 &-0.08 &-1.92 &2.66 (3.55)\\
        MnPS$_3$ &-1.54 &-0.14 &-0.36 &0.0086 (0.0108)\\
        \hline
    \end{tabular}
    \caption{In-plane exchange and anisotropy constants of bulk CrI$_3$ \cite{Chen-2021-cri3} and MPS$_3$ (M=Ni, Fe, Mn) \cite{Wildes2018} taken from INS experiments ($\mathrm{K}^{xx}=\mathrm{K}^{yy}=0$). All numbers are meV. The anisotropy constants are calculated from the spin-wave dispersion by fitting to either RPA or HP/RPA+CD (shown in brackets).}
    \label{table:1-J_n}
\end{table}

The Néel temperatures calculated using different approaches using INS data are tabulated in Tab. \ref{table:Tc}.
As we are considering the monolayer limit, we have ignored the reported interlayer exchange from Ref. \cite{Wildes2015}. The interlayer exchange constants are two orders of magnitude smaller that the intralayer ones and we are again assuming that the bulk interlayer exchange constants remain the same in the monolayer limit. 
We  observe that RPA+CD shows the closest agreement with experiments for NiPS$_3$. Since MnPS$_3$ has very weak anisotropy, the CD correction does not make a significant difference compared to RPA, and the latter shows slightly closer agreement with the experimental value. The parameters reported in Ref. \cite{Wildes2018} describe the anisotropy energy of MnPS$_3$ exclusively through the single-ion contribution, while the dipolar contribution, reported to have a comparable magnitude ($\Delta E^{xz}_\textrm{dip}\approx$ 54 $\mu$eV/site) \cite{Kim2021-mps3}, was not included in the analysis.
A consistent inclusion of the dipolar contribution requires the reanalysis and refitting of the experimental Heisenberg parameters of MnPS$_3$ to a model that includes dipolar interactions, but we have not attempted it here. It is expected to yield a small correction to the magnon energies and is not expected to influence the critical temperature significantly \cite{Wildes1998}. 
In the case of FePS$_3$, the RPA/RPA+CD deviates significantly from the experimental value although RPA+CD yields a somewhat better agreement than RPA. The HP method, while giving the closest agreement with experiment, generally exhibits a pathological behavior with discontinuous magnetization near the Néel temperature \cite{BeyondRPA2025,Helimag3d2025}, and therefore cannot be regarded as a reliable result in our opinion. 
It has previously been shown that both RPA, RPA+CD and HP severely overestimate critical temperatures when the anisotropy becomes large and all methods are thus expected to be inaccurate for FePS$_3$, which have $\mathrm{K}^{zz}/J_1\sim 1$ \cite{Torelli2019,BeyondRPA2025}. Moreover, the magnetic anisotropy in the INS data was solely fitted with the quadratic single-ion anisotropy term in the Hamiltonian. However, first principle calculations reveal that FePS$_3$ exhibits a fully unquenched orbital moment aligned with the out-of-plane direction \cite{Kim2021-mps3, ovesen_giant_2025} and the anisotropy is probably better described by an $\mathbf{L}\cdot\mathbf{S}$ term in the Hamiltonian (with fixed $\mathbf{L}$), which may give rise to a different behavior compared to single-ion anisotropy. 
Such a term can of course easily be included in the Heisenberg model and gives rise to effects that are hard to distinguish from single-ion anisotropy. However, at elevated temperatures, the orbital magnetization itself gets reduced by thermal (longitudinal) fluctuations, which cannot be described in the context of the Heisenberg model. We thus believe that an accurate description of the thermodynamics of FePS$_3$ requires a treatment that goes beyond the Heisenberg model.

\begin{figure}
    \includegraphics[scale=0.3]{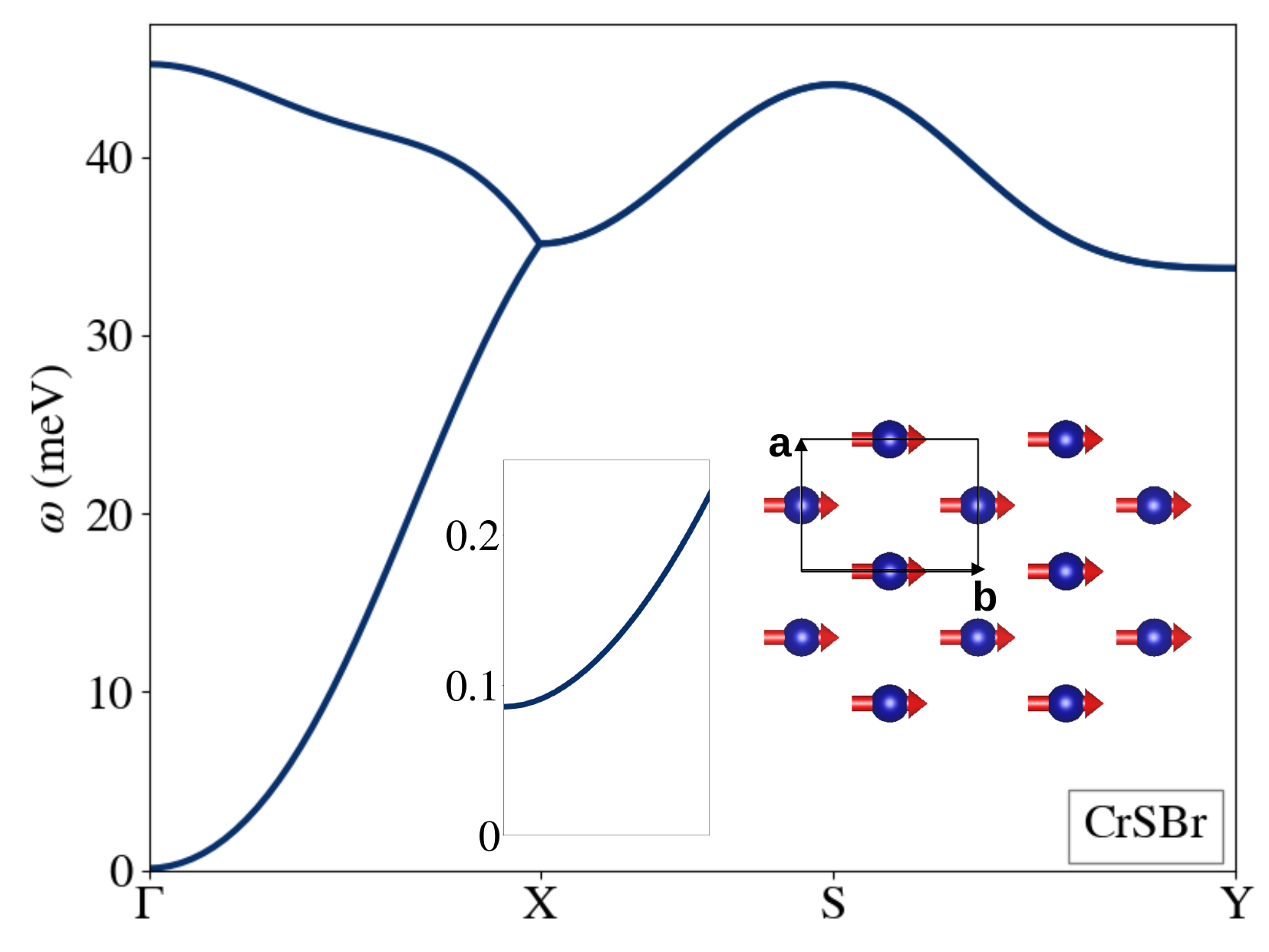}
    \label{fig:CrSBr}
    \caption{The magnon dispersion of CrSBr calculated from the experimental parameters listed in tab. \ref{table:2-J_n}. The inset shows the real space arrangement of magnetic atoms.}
\end{figure}

\subsection{CrSBr}
CrSBr is a layered van der Waals bonded anti-ferromagnetic semiconductor, in which the individual ferromagnetic layers are coupled anti-ferromagnetically \cite{Katscher1966-crsbr, Beck1990-crsbr, Goser1990-crsbr}. The ferromagnetic monolayer has been exfoliated from bulk and exhibits a remarkably high Curie temperature of 146 K \cite{Lee2021-CrSBr, Telfor2020-crsbr,Ziebel2024-CrSBr,Lopez2022-crsbr}. The Cr$^{3+}$ magnetic ions form a rectangular lattice with a ground state spin of $S=3/2$. In addition, it is notable for exhibiting triaxial magnetic anisotropy, with distinct easy, intermediate, and hard axes \cite{Bae2022-CrSBr-K,Yang2021-crsbr-dft}. 

\begin{table}[t]
    \centering
    \begin{tabular}{|p{1.5cm}|p{2.7cm}|}
        \hline
        \multicolumn{2}{|c|}{CrSBr} \\
        \hline
        $J_1$ & 1.9 \\
        $J_2$ & 3.38\\
        $J_3$ & 1.67 \\
        $J_4$ & 0.09 \\
        $J_5$ & 0.09 \\
        $J_7$ & -0.37 \\
        $J_8$ & 0.29 \\
        $\mathrm{K}^{xx}_a$ & 0.044/0.066/0.0528\\
        $\mathrm{K}^{yy}_a$ & 0.058/0.087/0.0804\\
        $\mathrm{K}^{zz}_a$ & 0.0/0.0/0.0\\
        \hline
    \end{tabular}
    \caption{In-plane exchange constants (in meV) of bulk CrSBr taken from Ref. \cite{Scheie2022}. $J_6$ is omitted as it corresponds to interlayer exchange in the layered compound. The anisotropy parameters (in meV) are taken from spectroscopic measurements \cite{Bae2022-CrSBr-K} and are converted (using isotropic invariance) such that the hard axis anisotropy component ($\mathrm{K}_a^{zz}$) is zero. The values are listed for RPA/RPA+CD/HP, which are obtained by fitting the predicted dispersion relations to the experimental one.}
    \label{table:2-J_n}
\end{table}

The Heisenberg parameters of CrSBr used in this work are compiled from two experimental references. The isotropic exchange interactions of bulk CrSBr are taken from Ref. \cite{Scheie2022} where the anisotropy was reported too weak to be measured. Since anisotropy is a crucial parameter for magnetic order, we use the values reported in Ref. \cite{Bae2022-CrSBr-K} where the hard and intermediate anisotropy parameters are obtained from anti-ferromagnetic resonance experiments. In CrSBr, the hard axis is along the $c$-axis, the intermediate is the $a$-axis and the  easy axis is the $b$-axis. The point group of the monolayer is $mmm$ and the mirror symmetries enforce all off-diagonal elements of the single-ion anisotropy tensors to vanish. We will again use the isotropic invariance of Eq. \eqref{eq:heisenberg-model_sia} to set the hard axis components $\mathrm{K}_a^{zz}$ to zero and we convert the values stated in Ref. \cite{Bae2022-CrSBr-K} to our conventions with a full characterization given by the components $\mathrm{K}_a^{xx}$ and $\mathrm{K}_a^{yy}$. 

The calculated magnon dispersion of CrSBr is shown in fig. \ref{fig:CrSBr} and the exchange and anisotropy parameters used in the calculations are tabulated in tab. \ref{table:2-J_n}.
The interlayer interactions are ignored in the calculations as they were reported to be negligible \cite{Scheie2022}. 
Unlike the uniaxial case, the HP and RPA+CD magnon gap differ at zero temperature due to the different in-plane components of magnetic anisotropy. Hence, the components of $\mathrm{K}_a$ have to be properly refitted for HP and RPA+CD to ensure that the same magnon dispersion is obtained for all three methods. The magnon gap induced by triaxial anisotropy is two orders of magnitude smaller than the magnon band width, yet it plays a crucial role in stabilizing the magnetic order in CrSBr. The Curie temperatures calculated using different approaches are summarized in tab. \ref{table:Tc}, and again RPA+CD shows the best agreement with the experimental value.

\begin{table}[t]
    \centering
    \begin{tabular}{ |p{1.2cm}|p{1cm}|p{1cm}|p{1cm}|p{1.5cm}|p{1.2cm}|}
        \hline
        &MC & HP & RPA & RPA+CD& Expt \\
        \hline
        CrI$_3$ & 15 & 20 & 52 & 47 & 45\\
        NiPS$_3$ & 85 &100 & 167 & 157 & 155*\\
        MnPS$_3$ &55 &44 & 67 & 64& 78*\\
        FePS$_3$ &90 & 118 & 266& 215 &120*\\
        CrSBr & 56 & 91 & 151& 150 &146\\
        \hline
    \end{tabular}
    \caption{Calculated critical temperatures (in K) of the materials considered in this work. The experimental values marked with a star are critical temperatures in the bulk compounds, while the remaining ones (including all calculations) are for monolayers. 
    The MC simulations are performed with a $30\times30$ super cell with 500,000 iteration steps.
    The HP and Green's function calculations are performed self-consistently by iteratively converging the magnetization at each temperature step.
    }
    \label{table:Tc}
\end{table}

\section{Conclusion}
We have extended the theoretical approaches for calculating critical temperatures of single-$Q$ 2D magnetic materials - emphasizing the role of single-ion anisotropy. In the HP approach, normal ordering of the fourth order bosonic operators yield second-order corrections to the zero temperature magnon energies. These corrections capture the correct behavior of uniaxial single-ion anisotropy for ferromagnets, but not for anti-ferromagnets or for general systems with triaxial anisotropy. 
Moreover, the magnetization calculated at finite temperatures in HP is know to exhibit a pathological discontinuity, which makes self-consistency loops hard to converge \cite{Helimag3d2025}. 
We then studied the RPA decoupling scheme applied to the single ion anisotropy terms and it was shown that RPA 
fails to reproduce the correct spin-wave gap - in particular the irrelevance of single-ion anisotropy for $S=1/2$. 
This comprises a severe flaw of the method when applied to systems with single-ion anisotropy. 
Finally, we extended Callen's decoupling scheme to include the single-ion anisotropy tensor of any single-$Q$ system. In contrast to RPA, CD captures the correct spin-wave gap and irrelevance of single-ion anisotropy for $S=1/2$ ferromagnets. Moreover, in contrast to HP,  
its accuracy is not restricted to uniaxial cases and we therefore expect CD to provide the best description of single-ion anisotropy.
However, for isotropic models with $S=1/2$ it has previously been shown that RPA provides more accurate estimates of critical temperatures compared to Callen decoupling \cite{Frobich2006}.
We thus adopt a hybrid RPA+CD approach, using Callen’s decoupling for the single-ion anisotropy and RPA for the remaining interactions. This combination has also been shown to yield the best agreement with  quantum Monte Carlo results for ferromagnets \cite{Frobich2006}.

We applied the three methods to a set of two-dimensional materials with exchange parameters taken from experiments and compared the calculated critical temperatures to experimental values. 
In comparison with experimental critical temperatures, we found that RPA+CD gives the best performance, except for MnPS$_3$ and FePS$_3$. For MnPS$_3$, which is nearly isotropic with only a weak uniaxial anisotropy, the application of the CD correction yields a negligible deviation from the RPA results. In addition, incorporating the dipolar contribution into the fit of the magnon anisotropy from INS data would likely improve the precision of the results. In the case of FePS$_3$, the the Fe atoms carry a large orbital moment, which is pinned to the lattice and strongly coupled to the spin. This indicates that a simple single-ion anisotropy tensor is likely insufficient to fully describe the magnetic anisotropy. In any case, the RPA+CD approach was found perform better than RPA, while HP yields excellent agreement with the experimental value. However, we believe that this is a fortuitous result originating in error cancellation within that method. 
The classical MC simulations underestimate the critical temperatures as observed for bulk materials. Although such results may be improved by simple quantum corrections \cite{Helimag3d2025}, the RPA+CD provides a more physically sound and fully quantum mechanical alternative that does not require extensive numerical simulations.

In summary, we presented a comprehensive analysis of the role of single-ion anisotropy in calculating thermally excited magnon energies and critical temperatures in two dimensional magnets. Among the methods considered, the RPA+CD approach showed the best agreement with experiments, establishing it as the best method for predicting critical temperatures in low-dimensional systems. This approach comprises a simple and versatile method that is easily applicable and is expected to significantly improve the accuracy of either classical Monte Carlo simulations or linear spin wave theory, which seems to be the most common approaches in the literature. 
Although no explicit benchmarking was performed for non-collinear states, all of these methods are directly applicable to such systems, provided they retain the single-$Q$ symmetry.
The framework can also be used to analyze the influence of complex spin interactions, such as anisotropic exchange, Dzyaloshinskii-Moriya interaction, dipole-dipole interactions and higher order terms like biquadratic exchange. As such, we believe that the RPA+CD approach provides the most reliable route for understanding and predicting thermal properties of 2D materials.

\appendix

\section{Exchange contribution to the HP Hamiltonian}\label{app:exchange_HP}
The exchange contribution to the constant in Eq. \eqref{eq:heisenberg-hp} is
\begin{align}
    \mathcal{E}_0 = -\frac{1}{2}\sum_{abij} \mathbf{v}^T_a \textrm{J}^\prime_{abij}\mathbf{v}_b( S_a S_b + \frac{S_a+S_b}{2}),
\end{align}
and the exchange contribution to the quadratic Hamiltonian is
\begin{align}
    \mathrm{H}^\mathrm{HP}_{\mathbf{q}} &=\mathrm{H}_{\mathbf{q}}^0 + \mathrm{H}_{\mathbf{q}}^1 ,\\
    \mathrm{H}_{\mathbf{q}}^0 &= -\frac{1}{2}
    \begin{bmatrix}
        \mathrm{A}^0_{ab\mathbf{q}} - \mathrm{C}^0_{ab} & \mathrm{B}^0_{ab\mathbf{q}} \\
        \mathrm{B}^{0 \dagger}_{ab\mathbf{q}}& \mathrm{A}^{0 *}_{ab\mathbf{-q}} - \mathrm{C}^0_{ab}
    \end{bmatrix}\label{eq:H0_q},\\
    \mathrm{H}_{\mathbf{q}}^1 &= \frac{1}{2}
    \begin{bmatrix}
          \mathrm{A}^1_{ab\mathbf{q}}- \mathrm{C}^1_{ab\mathbf{q}} & \mathrm{B}^1_{ab\mathbf{q}} \\
        \mathrm{B}^{1 \dagger}_{ab\mathbf{q}} & \mathrm{A}^{1 *}_{ab-\mathbf{q}} - \mathrm{C}^{1 *}_{ab\mathbf{q}}.
    \end{bmatrix} \label{eq:H1_q},
\end{align}
with
\begin{align}
    \mathrm{A}^0_{ab\mathbf{q}} &= \frac{\sqrt{S_aS_b}}{2}  \mathbf{u}_a^T\mathrm{J}^{\prime}_{ab\mathbf{q}} \mathbf{u}_b^* \label{eq:HP-comp1},\\
    \mathrm{B}^0_{ab\mathbf{q}} &=\frac{\sqrt{S_aS_b}}{2}  \mathbf{u}_a^T\mathrm{J}^{\prime}_{ab\mathbf{q}} \mathbf{u}_b, \\
    \mathrm{C}^0_{ab} &= \delta_{ab} \sum_c S_c \mathbf{v}^T_a \mathrm{J}^\prime_{ac\mathbf{0}} \mathbf{v}_c,
\label{eq:HP0-compL}
\end{align}
and
\begin{align}
    &\mathrm{A}^1_{ab\mathbf{q}}  = \frac{1}{N} \sum_\mathbf{q^\prime}\bigg( \mathbf{u}^T_a\mathrm{J}^\prime_{ab\mathbf{q}} \mathbf{u}^*_b \frac{n_{aa\mathbf{q^\prime}} + n_{bb\mathbf{q^\prime}}}{4} \notag\\&\quad +  \delta_{ab} \sum_{c} \frac{\mathbf{u}^T_c\mathrm{J}^\prime_{ca\mathbf{q^\prime}}\mathbf{u}^*_a n_{ca\mathbf{q^\prime}} + \mathbf{u}^T_a\mathrm{J}^\prime_{ac\mathbf{q^\prime}} \mathbf{u}^*_c n_{ac\mathbf{q^\prime}} }{4}\bigg), \\
    &\mathrm{B}^1_{ab\mathbf{q}} = \frac{1}{N} \sum_\mathbf{q^\prime} \bigg( \mathbf{u}^T_a\mathrm{J}^\prime_{ab\mathbf{q}} \mathbf{u}_b \frac{n_{aa\mathbf{q^\prime}} + n_{bb\mathbf{q^\prime}}}{4} \notag\\&\quad+  \delta_{ab} \sum_{c} \frac{\mathbf{u}^T_a\mathrm{J}^\prime_{ac\mathbf{q^\prime}} \mathbf{u}_c n_{ac\mathbf{q^\prime}} + \mathbf{u}^T_c\mathrm{J}^\prime_{ac\mathbf{q^\prime}} \mathbf{u}_a n_{ac\mathbf{q^\prime}}}{8}\bigg), \\
    &\mathrm{C}^1_{ab\mathbf{q}} = \frac{1}{N}  \sum_\mathbf{q^\prime} \bigg(\mathbf{v}^T_a \mathrm{J}^\prime_{ab\mathbf{q-q^\prime}}\mathbf{v}_b n_{ba\mathbf{q^\prime}}\notag\\&\qquad\qquad +\delta_{ab} \sum_c n_{cc\mathbf{q^\prime}} \mathbf{v}^T_a \mathrm{J}^\prime_{ac\mathbf{0}}\mathbf{v}_c\bigg).\label{eq:HP1-compL}
\end{align}
The linear and cubic order bosonic terms do not contribute to the dispersion as they can be removed by a linear transformation.
Note that the expressions here are slightly different from the ones presented in Ref. \cite{Helimag3d2025} because we have separated the single-ion anisotropy tensor from $\mathrm{J}_{ab\mathbf{q}}$. Here the $q$-space exchange interactions are defined as
\begin{align}
    \mathrm{J}^\prime_{ab\mathbf{q}} &= \sum_i \mathrm{J}_{ab0i}  e^{i\mathbf{q}\cdot \mathbf{r}_i}\label{eq:Jprime_q},\\
    \mathrm{J}^\prime_{abij} &= \mathrm{R}^T_i \mathrm{J}_{abij}\mathrm{R}_j.
\end{align}
and the site occupation matrix element is given by
\begin{align}
   n_{ab\mathbf{q}} &= \sum_n [\mathrm{T}_\mathbf{q}]_{a+N_a, n+N_a} [\mathrm{T}^*_\mathbf{q}]_{n+N_a,b+N_a} n_\mathrm{B}(\omega_{n,\mathbf{q}}), \label{eq:n_abq}
\end{align}
where $n_\mathrm{B}(\omega_{n,\mathbf{q}})$ is the Bose distribution and $\mathrm{T}_\mathbf{q}$ is the canonical matrix that transforms $\mathbf{a}_{\mathbf{q}}$ to a basis that (para)diagonalizes the Hamiltonian \eqref{eq:heisenberg-hp} \cite{Toth2015,Helimag3d2025}
\begin{align}\label{eq:gen-BogVal}
     \mathbf{a}_\mathbf{q}&= \mathrm{T}_\mathbf{q} \boldsymbol{\alpha}_\mathbf{q}\\
    \boldsymbol{\alpha}_{\mathbf{q}} &= 
    \begin{bmatrix}
       \alpha_{n,\mathbf{q}} & \alpha^{\dagger}_{n,\mathbf{-q}}
    \end{bmatrix}^T  n\in [1,N_a]\\
    (\mathcal{H}-E_0)\alpha^{\dagger}_{n,\mathbf{q}}|0\rangle&=\omega_{n,\mathbf{q}}\alpha^{\dagger}_{n,\mathbf{q}}|0\rangle,
\end{align}
where $E_0=\mathcal{E}_0+\mathcal{E}^0_\mathrm{K}$ is the ground state energy. In this basis, the magnon number can be evaluated as
\begin{align}\label{eq:PhiHP}
    \Phi_a =& \frac{1}{N}\sum_{\mathbf{q}} \big[\mathrm{T}_\mathbf{q} \langle \boldsymbol{\alpha}_{\mathbf{q}} \boldsymbol{\alpha}^\dagger_{\mathbf{q}} \rangle \mathrm{T}^{\dagger}_\mathbf{q} \big]_{a + N_a, a + N_a}, \\
    \langle \boldsymbol{\alpha}_{\mathbf{q}} \boldsymbol{\alpha}^\dagger_{\mathbf{q}} \rangle &=
   \begin{bmatrix}
        1+n_\mathrm{B}(\omega_{n,\mathbf{q}}) &  0 \\
        0 & n_\mathrm{B}(\omega_{n,\mathbf{q}})
   \end{bmatrix}.
\end{align}
The procedure is described in more detail in Ref. \cite{Helimag3d2025} and Ref. \cite{helimagthesis}. 

\section{Exchange contribution to the RPA Hamiltonian}\label{app:RPA}
The exchange contribution to the RPA Hamiltonian \eqref{eq:eqnofmotion} is given by
\begin{align}
    \mathrm{H}^\mathrm{RPA}_\mathbf{q} &=
    \begin{bmatrix}
        -\mathrm{A}^\mathrm{RPA}_{ab\mathbf{q}} +  \mathrm{C}^\mathrm{RPA}_{ab} & -\mathrm{B}^\mathrm{RPA}_{ab\mathbf{q}} \\
         (\mathrm{B}^\mathrm{RPA}_{ab\mathbf{q}})^\dagger & (\mathrm{A}^\mathrm{RPA}_{ab\mathbf{-q}})^* - \mathrm{C}_{ab} 
    \end{bmatrix}\label{eq:H_rpa}, \\
    \mathrm{A}^\mathrm{RPA}_{ab\mathbf{q}} &= \frac{\langle S^{\dprime z}_a \rangle}{2}\mathbf{u}^{T}_a \mathrm{J}^{\prime}_{ab\mathbf{q}}\mathbf{u}^*_b,\label{eq:A-rpa}\\
    \mathrm{B}^\mathrm{RPA}_{ab\mathbf{q}} &= \frac{\langle S^{\dprime z}_a \rangle}{2}  \mathbf{u}^{T}_a \mathrm{J}^{\prime}_{ab\mathbf{q}}\mathbf{u}_b,\\
    \mathrm{C}^\mathrm{RPA}_{ab} &= \delta_{ab}\sum_c \langle S^{\dprime z}_c \rangle  \mathbf{v}^T_a \mathrm{J}^{\prime}_{ac\mathbf{0}}\mathbf{v}_c,\label{eq:C-rpa}
\end{align}
with $\mathrm{J}^{\prime}_{ab\mathbf{q}}$ defined as in Eq. \eqref{eq:Jprime_q}. This exchange Hamiltonian is used for both the RPA and RPA+CD approaches \cite{Helimag3d2025,helimagthesis}.

\section{Magnon dispersion at zero temperature for Bravais lattices}
In the case of Bravais lattices, the HP Hamiltonian at $T=0$ can be expressed as
\begin{align}
    \mathrm{H}^\mathrm{HP}_\mathbf{q} &= -\frac{S}{2}
    \begin{bmatrix}
        X^\mathrm{HP}_\mathbf{q} & Y^\mathrm{HP}_\mathbf{q} \\
        (Y^\mathrm{HP}_\mathbf{q})^* & X^\mathrm{HP}_\mathbf{q}
    \end{bmatrix},\label{eq:Hhp-bravais}
\end{align}
and the magnon energies can be obtained by the canonical transformation of this Hamiltonian:
\begin{align}
    {\mathrm{T}_\mathbf{q}^{ \dagger}}\mathrm{H}^\mathrm{HP}_{\mathbf{q}}\mathrm{T}_\mathbf{q} &=\frac{1}{2}
    \begin{bmatrix}
        \omega^\mathrm{HP}_\mathbf{q} & 0 \\
        0 & \omega^\mathrm{HP}_\mathbf{-q}
    \end{bmatrix}, \\
    \omega^\mathrm{HP}_\mathbf{q} &= S\sqrt{{X^\mathrm{HP}_\mathbf{q}}^2 - |Y^\mathrm{HP}_\mathbf{q}|^2}.
\end{align}
The details of the required paradiagonalization and the construction of the transformation $\mathrm{T}_\mathbf{q}$ can be found in Refs. \cite{Helimag3d2025,Toth2015,COLPA1978327}. 

For the case of RPA, the magnon energies are easily obtained by diagonalizing the $T=0$ Hamiltonian
\begin{align}
        \mathrm{H}^\mathrm{RPA}_\mathbf{q} &= S^{(0)}
        \begin{bmatrix}
        -X^\mathrm{RPA}_\mathbf{q} & -Y^\mathrm{RPA}_\mathbf{q} \\
        (Y^\mathrm{RPA}_\mathbf{q})^* & X^\mathrm{RPA}_\mathbf{q}
    \end{bmatrix},\label{eq:Hrpa-bravais}\\
    \omega^\mathrm{RPA}_\mathbf{q} &= S^{(0)}\sqrt{{X^\mathrm{RPA}_\mathbf{q}}^2 - |Y^\mathrm{RPA}_\mathbf{q}|^2}.
\end{align}
If we exclude the contributions form Eq. \eqref{eq:Hhp_2} to $X^\mathrm{HP}_\mathbf{q}$ and $Y^\mathrm{HP}_\mathbf{q}$, the terms of order $\frac{1}{S}$ in Eqs. \eqref{eq.xhp_q}-\eqref{eq:yhp_q} are not present. In this case the dispersion relations predicted by HP and RPA at $T=0$ are identical except the latter is renormalized by the saturation magnetization $S^{(0)}$ rather than the spin eigenvalue $S$.

Finally, for the case of RPA+CD, the magnon dispersion is given by
\begin{align}
    \omega^\mathrm{RPA+CD}_\mathbf{q} &= S^{(0)}\sqrt{{X^\mathrm{RPA+CD}_\mathbf{q}}^2 - |Y^\mathrm{RPA+CD}_\mathbf{q}|^2},\\
    X^\mathrm{RPA+CD}_\mathbf{q} &= \frac{\mathbf{u}^T\mathrm{J}^\prime_\mathbf{q}\mathbf{u}^*}{2} - \mathbf{v}^T\mathrm{J}^\prime_\mathbf{q} \mathbf{v} \notag\\&+ \big(\mathbf{u}^T\mathrm{K}\mathbf{u}^* - 2\mathbf{v}^T\mathrm{K}\mathbf{v} \big)\Upsilon_\mathrm{CD} , \\
    Y^\mathrm{RPA+CD}_\mathbf{q} &= \frac{\mathbf{u}^T\mathrm{J}^\prime_\mathbf{q}\mathbf{u}}{2} + \mathbf{u}^T\mathrm{K}\mathbf{u}\Upsilon_\mathrm{CD}, \\
    \Upsilon_\mathrm{CD} &= 1- \frac{(S-\Phi^0)(1+2\Phi^0)}{2S^2}.
\end{align}

\bibliography{references}
\end{document}